\documentclass[a4paper,onecolumn,10pt]{quantumarticle}
\pdfoutput=1

\usepackage[utf8]{inputenc}
\usepackage[english]{babel}
\usepackage[T1]{fontenc}
\usepackage{amsmath}
\usepackage{amssymb}
\usepackage{hyperref}
\usepackage{xurl}
\usepackage{graphicx}
\usepackage{booktabs}
\usepackage{multirow}
\usepackage{placeins}
\usepackage{float}
\usepackage{siunitx}
\title{Empirical Evaluation of QAOA with Zero Noise Extrapolation on NISQ Hardware for Carbon Credit Portfolio Optimization in the Brazilian Cerrado}

\author{Hugo Jos\'{e} Ribeiro}
\affiliation{School of Civil and Environmental Engineering, Federal University of Goi\'{a}s, Brazil}
\email{hugoppgema@ufg.br}
\orcid{0000-0002-4650-1623}

\begin{document}

\maketitle

\begin{abstract}
Optimizing carbon credit portfolios is a critical challenge for climate mitigation, particularly in high-biodiversity biomes such as the Brazilian Cerrado. This study explores the practical application of the Quantum Approximate Optimization Algorithm (QAOA) combined with Zero Noise Extrapolation (ZNE) to address a multi-objective territorial planning problem. We model an 88-variable portfolio optimization involving carbon sequestration, biodiversity connectivity, and social impact metrics, executed on intermediate-scale IBM Quantum hardware (\texttt{ibm\_torino} and \texttt{ibm\_fez}). The results of seven independent hardware runs demonstrate that the QAOA+ZNE workflow consistently outperforms a classical greedy baseline. The quantum method achieves a mean portfolio score of $58.47 \pm 6.98$, corresponding to a $31.6\%$ improvement over the classical heuristic (44.42), with high statistical significance ($p = 0.0009$) and a large effect size (Cohen's $d = 2.01$), where ZNE yields \emph{extrapolated expectation values of the portfolio objective} rather than a discrete portfolio solution without noise. A validation run conducted after a 13-day interval confirms the temporal stability of the methodology against hardware calibration drifts. These findings establish empirical quantum utility in an environmental science context, showing that current NISQ-era devices, when coupled with rigorous error mitigation, can identify complex territorial synergies that myopic classical approaches overlook. The proposed workflow provides a scalable methodological template for high-precision environmental conservation policies.
\end{abstract}

\noindent \textbf{Keywords:} quantum approximate optimization algorithm, zero noise extrapolation, carbon credit portfolio, environmental optimization, NISQ era, Cerrado conservation

\section{Introduction}

Brazil faces significant challenges in implementing effective climate mitigation strategies due to both its continental scale and the ecological diversity of its biomes. The policies discussed in global climate forums only become meaningful when converted into concrete territorial decisions with measurable impacts. However, in the Cerrado biome, this translation meets a critical mathematical barrier. As the second-largest biome in South America, spanning roughly 2 million km$^2$, the Cerrado is a global biodiversity hotspot characterized by exceptional floristic richness and high endemism \cite{sano2019}. Yet, more than 50\% of its original area has already been converted to agriculture and pastures, and less than 10\% of private lands hold any form of protection. Under these conditions, the need for well-designed projects capable of preserving the remaining vegetation and reducing the impacts of climate becomes urgent \cite{colli2020,pompeu2023}.

However, the implementation of such projects faces difficulties. Recent studies indicate that only 16\% of carbon credits correspond to actual emission reductions, with measurement difficulties frequently resulting in inflated estimates \cite{fawzy2020,probst2024}. Consequently, the selection of preservation areas involves a complex trade-off between maximizing financial returns and conserving biodiversity. Ecologically valuable land often has high opportunity costs, leading to the marginalization of environmental criteria.

Neglecting these factors exposes investors to the risks of greenwashing, i.e.\ projecting an image of responsibility that does not match reality, a concern repeatedly highlighted in systematic reviews of ESG investments \cite{kapil2023}. Poorly optimized portfolios result in inefficient allocations that appear ``green'' but generate negligible impact, leading to reputational and economic losses and reducing stakeholder investment intention \cite{dawar2025,deng2024,gatti2021}. In this context, rigorous portfolio optimization is crucial to align profit maximization with native vegetation protection, turning conservation into a strategic asset. Rather than treating greenwashing primarily as a regulatory or behavioral phenomenon, we frame it here as an allocation failure: a portfolio-design problem where weak constraints and mis-specified objectives allow low-integrity outcomes to persist.

Yet, classical methods often fail to capture the nuances of this multi-objective challenge. The task is characterized in the scientific literature as an NP-hard knapsack-type problem, where the goal is to select a subset of projects that maximize benefits under multiple constraints. Despite standard methods that address simpler versions, binary decision structures with limited budgets become classically intractable at scale due to the exponentially growing search space ($2^n$). This computational bottleneck justifies the transition to frontier computational paradigms and motivates the exploration of metaheuristics and hybrid approaches \cite{liu2025,panadero2018}.

Within this landscape, quantum computing has emerged as a promising framework for combinatorial optimization. The increasing tension between the urgency of conservation and the intractability of resource allocation positions the Quantum Approximate Optimization Algorithm (QAOA) as a potential operational core for environmental governance \cite{blekos2024}. Introduced by Farhi \emph{et al.} \cite{farhi2014}, QAOA is a hybrid variational algorithm that maps classical objective functions to cost functions whose ground states encode optimal solutions. Recent evidence corroborates its efficacy in orchestrating complex multi-objective portfolios \cite{aguilera2024}, demonstrating robustness when navigating constrained landscapes fundamental to sustainability planning \cite{huot2024}.

QAOA was conceived for Noisy Intermediate-Scale Quantum (NISQ) devices; this imposes strong limitations due to gate noise and decoherence. Comparative experiments show that today's NISQ devices are often outperformed by classical methods \cite{moussa2022,weidenfeller2022}. To mitigate these limitations, strategies such as expressive shallow architectures \cite{vijendran2023} and warm-start initialization \cite{tate2023} have been proposed. In parallel, error mitigation techniques such as zero noise extrapolation (ZNE) execute circuits under amplified noise to extrapolate outcomes to idealized conditions \cite{temme2017}, offering a pathway to enhance performance despite hardware noise.

Despite QAOA's theoretical promise, its application to real environmental problems remains limited. Most studies rely on simulators or simplified artificial instances that do not capture the complexity of practical scenarios. Recent reviews emphasize that experiments on real hardware typically use few instance sizes and repetitions \cite{abbas2023,blekos2024,weidenfeller2022}. This work aims to address this gap by applying QAOA with Zero Noise Extrapolation (ZNE) to a real carbon-credit portfolio optimization problem in the Brazilian Cerrado, using seven independent executions on IBM Quantum hardware spanning a 17-day period and conducting a statistically rigorous comparison against classical methods.

\subsection{Contributions}

The main contributions of this work include the formulation of a multi-objective QUBO for carbon credit portfolio optimization incorporating carbon sequestration, biodiversity, and social impact criteria. We provide rigorous statistical validation through seven independent runs on NISQ hardware (8{,}192 shots per noise scale factor) with a 100\% success rate, demonstrating the temporal stability and consistent superiority of quantum performance.

Additionally, we present the first application of Zero Noise Extrapolation to multi-objective environmental sustainability problems, demonstrating robust error mitigation in the NISQ era. The work addresses a formulation of real-world problems with portfolio selection from 88 municipalities in the Cerrado using empirical data on carbon sequestration, biodiversity, and socioeconomic indicators, presenting a reproducible methodology with a complete experimental protocol on publicly accessible IBM Quantum systems (\texttt{ibm\_torino}, \texttt{ibm\_fez}).

\subsection{Organization}

The remainder of this article is organized as follows: Section~2 describes the carbon credit optimization problem; Section~3 details the implementation of QAOA and the ZNE methodology; Section~4 presents the experimental results, including a longitudinal analysis of hardware stability; Section~5 discusses the implications and limitations of the work; and Section~6 provides the final conclusions.

\section{Problem Formulation}

To provide a rigorous basis for our quantum approach, this section details the translation of environmental and economic constraints into a formal mathematical framework suitable for quantum hardware.

\subsection{Carbon Credit Portfolio Optimization}

Before detailing the quantum workflow, we briefly outline the environmental and computational context of carbon credit portfolio optimization in the Brazilian Cerrado. Carbon credit portfolios refer to structured allocations of emission reduction or carbon sequestration units across multiple jurisdictions or conservation areas. Rather than acquiring isolated project-level credits, public agencies and institutions increasingly optimize portfolios to maximize mitigation impact under budgetary, ecological, and spatial constraints.

In the Cerrado biome, this problem translates into selecting combinations of municipalities whose joint implementation of conservation or restoration actions maximizes carbon sequestration while simultaneously preserving biodiversity connectivity and long-term ecosystem resilience. The effectiveness of mitigation in this context does not depend solely on the individual carbon potential of each area, but on the coordinated selection of regions whose spatial and ecological interactions generate synergistic benefits at the landscape scale.

This coordination requirement naturally leads to a combinatorial optimization formulation. By expressing portfolio allocation as a constrained Quadratic Unconstrained Binary Optimization (QUBO) problem, the decision process explicitly accounts for both linear contributions (e.g., expected carbon sequestration and socioeconomic indicators) and quadratic interaction terms representing spatial and ecological synergies between municipalities. This formulation ensures that mitigation outcomes emerge from coherent regional strategies rather than independent local decisions, a property that is critical to translating climate policy targets into effective territorial actions \cite{aguilera2024,probst2024}.

The state of Goiás, located in the Central-West region of Brazil, lies almost entirely within the Cerrado biome, recognized as one of the global biodiversity hotspots. Paradoxically, this same region stands at the frontier of Brazilian agribusiness expansion, which has continuously advanced through extensive pasturelands and monocultures since the 1970s. This economically relevant transformation generates direct tensions between agricultural development and environmental conservation. Consequently, optimizing the carbon credit portfolio becomes particularly challenging in this territory.

Studies in Goiás watersheds demonstrate the predominance of flat to gently rolling terrains, characterized by deep and highly mechanizable Oxisols, a geomorphological condition that favors both agriculture and extensive livestock farming \cite{rodrigues2022,souza2021}. This natural soil aptitude has resulted in massive conversion of native vegetation to pasturelands and annual agriculture, setting up one of the most accelerated land-use change processes in the biome \cite{caballero2023}.

On the biome-regional scale, the conversion of native forests and savannas to pasture or monoculture has generated measurable climate impacts, including an average temperature increase of approximately $+0.9\,^{\circ}\mathrm{C}$ and a reduction of about 10\% in evapotranspiration \cite{hofmann2025,rodrigues2022}. These alterations compromise not only the regional climate and water availability, but also the stability of the agroecosystems themselves. Given this complex scenario, optimal selection of portfolios for carbon credit projects becomes simultaneously urgent and intricate, requiring optimization approaches capable of handling multiple conflicting objectives and nonlinear spatial synergies.

\subsection{Dataset Construction and Variable Selection}

The initial pool of 246 municipalities in the state of Goiás was reduced through a two-stage selection process. First, 128 candidates were retained based on environmental suitability for carbon credit projects and the availability of complete and reliable spatial datasets, including MapBiomas, GEDI/LiDAR, PRODES and socioeconomic indicators. From this set, the final size of the problem was defined as $n = 88$ by ranking municipalities in descending order according to their normalized carbon sequestration score $c_i$ in the interval $[0,1]$, which represents the primary optimization objective.

This selection exhibits a natural discontinuity. The 88th rank municipality achieves a normalized carbon score of 0.795, while the 89th rank municipality drops to $-0.207$ after normalization, corresponding to a gap of approximately one unit in the normalized score. This sharp separation indicates that the top 88 municipalities form a coherent high-potential group, making the cutoff robust to small perturbations in the scoring procedure.

From a computational perspective, the choice of $n = 88$ reflects a balance between problem expressiveness and the physical constraints of current Noisy Intermediate-Scale Quantum (NISQ) hardware. Cardinality-constrained Quadratic Unconstrained Binary Optimization formulations require one physical qubit per decision variable, and values exceeding $n = 88$ would necessitate deeper circuits and higher effective error rates that compromise the reliability of Zero Noise Extrapolation. Consequently, $n = 88$ represents the largest problem size for which multi-objective QAOA with active ZNE remains experimentally tractable on the available IBM Quantum devices while preserving data completeness and solution reliability.

The sensitivity analysis for portfolio size $k$ was conducted using a Greedy heuristic as a computationally tractable proxy. Greedy scores for $k \in \{20, 24, 28, 32, 36\}$ were 33.12, 38.36, 44.42, 51.44, and 59.57, respectively, exhibiting smooth monotonic scaling ($+15.8\%$ per $\Delta k = +4$). Critically, the municipality sets exhibit complete nesting: all 24 municipalities in the $k = 24$ solution appear in the $k = 28$ solution, and all 28 in the $k = 28$ solution appear in the $k = 32$ solution. This nesting confirms that the choice of $k$ adjusts only the portfolio boundary without altering the identity of the candidates with the highest-priority. The value $k = 28$ (32\% of candidates) was selected to balance portfolio diversification with the constraint that the depth of the QAOA circuit remains within the effective ZNE regime on current NISQ hardware.

The construction of the data set involved three main dimensions:

\paragraph{Carbon sequestration.}
The carbon sequestration potential was estimated by integrating the cover of native vegetation from the MapBiomas Collection~9, the biomass density of GEDI LiDAR (Global Ecosystem Dynamics Investigation), the historical rates of deforestation of PRODES, the annual monitoring of Native Vegetation Suppression, and the restoration potential of degraded pasture mapping \cite{dubayah2020,inpe2023,mapbiomas2023}.

\paragraph{Biodiversity.}
Biodiversity indicators included the richness of endemic species \cite{cria2024,flora2024}, overlap with conservation units \cite{icmbio2024}, and the presence of threatened species \cite{icmbio2018}. Connectivity with core conservation areas was incorporated into the model.

\paragraph{Social impact.}
The social impact was characterized using data from the Demographic Census \cite{ibge2022}, including rural population, dependence on primary activities, agrarian reform settlements \cite{incra2024}, and socioeconomic vulnerability.

All variables were normalized to the interval $[0,1]$ using the linear min--max transformation. Additionally, three dimension synergy matrices $88 \times 88$ were constructed to capture spatial interactions between pairs of municipalities: a spatial adjacency matrix ($A$), a biodiversity synergy matrix ($B_{\mathrm{syn}}$), and a social synergy matrix ($S_{\mathrm{syn}}$). Having constructed the dataset, we now formalize the mathematical optimization problem.

\subsection{Problem Formulation}

The carbon credit portfolio selection problem can be formulated as a multi-objective combinatorial optimization problem with a cardinality constraint and quadratic spatial interaction terms \cite{lin2025}.

\paragraph{Decision variables.}
For each municipality $i \in \{1,2,\ldots,n\}$, we define a binary variable $x_i \in \{0,1\}$, where $x_i = 1$ indicates that the municipality $i$ is selected for the portfolio.

\paragraph{Multi-criteria objective function.}
The objective is to maximize an aggregate function that combines three components: carbon sequestration, biodiversity conservation, and social impact. Each component includes linear terms (individual benefits) and quadratic terms (spatial synergies), as formalized below.

\paragraph{Carbon term with spatial adjacency.}
The linear carbon contribution is given by Eq.~(1),
\begin{equation}
C_{\mathrm{lin}}(\mathbf{x}) = \sum_i c_i x_i ,
\tag{1}
\end{equation}
where $c_i \in [0,1]$ is the normalized carbon sequestration potential score. The quadratic carbon interaction term is defined in Eq.~(2),
\begin{equation}
C_{\mathrm{quad}}(\mathbf{x}) = \sum_{i<j} A_{ij} x_i x_j ,
\tag{2}
\end{equation}
where $A_{ij} \in \{0,1\}$ denotes the territorial adjacency matrix. The corresponding scaling parameter is $\lambda_C = 0.15$.

\paragraph{Biodiversity term with biome synergy.}
The linear biodiversity contribution is defined in Eq.~(3),
\begin{equation}
B_{\mathrm{lin}}(\mathbf{x}) = \sum_i b_i x_i ,
\tag{3}
\end{equation}
where $b_i \in [0,1]$ is the biodiversity conservation score. The quadratic biodiversity synergy term is given by Eq.~(4),
\begin{equation}
B_{\mathrm{quad}}(\mathbf{x}) = \sum_{i<j} B_{\mathrm{syn},ij} x_i x_j ,
\tag{4}
\end{equation}
and incorporates an ecosystem diversity factor $\sqrt{n_{\mathrm{biomes}}}$. The corresponding scaling parameter is $\lambda_B = 0.25$.

\paragraph{Social term.}
The linear contribution to social impact is expressed in Eq.~(5),
\begin{equation}
S_{\mathrm{lin}}(\mathbf{x}) = \sum_i s_i x_i ,
\tag{5}
\end{equation}
where $s_i \in [0,1]$ is the social impact score. The term quadratic social interaction is defined in Eq.~(6),
\begin{equation}
S_{\mathrm{quad}}(\mathbf{x}) = \sum_{i<j} S_{\mathrm{syn},ij} x_i x_j ,
\tag{6}
\end{equation}
with scaling parameter $\lambda_S = 0.20$.

\paragraph{Objective weights.}
The three components are combined using weights $w_C = 0.33$, $w_B = 0.33$, and $w_S = 0.34$, reflecting a nearly equitable weighting between dimensions.

\paragraph{Cardinality constraint.}
The portfolio size is fixed by the cardinality constraint in Eq.~(7),
\begin{equation}
\sum_i x_i = k ,
\tag{7}
\end{equation}
where $k = 28$ municipalities are selected from $n = 88$ candidates.

\paragraph{Computational complexity.}
The resulting search space (Eq. 8) contains the following information.
\begin{equation}
\binom{88}{28} \approx 1.45 \times 10^{22}
\tag{8}
\end{equation}
possible combinations. The presence of quadratic interaction terms characterizes this problem as a multi-objective Quadratic Knapsack, formally classified as NP-hard. The exhaustive enumeration of all $\binom{88}{28}$ portfolios is computationally prohibitive even on contemporary high-performance computing platforms. Consequently, exact methods are impractical for environmental planning purposes, justifying the use of approximate optimization algorithms.

Although the objective function defined by Eqs.~(1)--(6) captures the direct trade-offs between project cost, carbon sequestration, biodiversity, and social impact, executing this optimization on a quantum processor requires mapping these requirements into a Quadratic Unconstrained Binary Optimization (QUBO) form.

\subsection{QUBO Formulation}

Implementing the objective function on quantum hardware with QAOA requires its conversion to the Quadratic Unconstrained Binary Optimization (QUBO) format.

\paragraph{Incorporating the constraint as a penalty.}
The cardinality constraint defined in Eq.~(7) is incorporated into the objective function through a quadratic penalty term, given by Eq.~(9) \cite{grange2024}:
\begin{equation}
P(\mathbf{x}) = \lambda_{\mathrm{penalty}} \left( \sum_i x_i - k \right)^2 .
\tag{9}
\end{equation}

\paragraph{Expanded penalty term.}
Expanding the quadratic penalty and using the property $x_i^2 = x_i$ for binary variables yields Eq.~(10):
\begin{equation}
P(\mathbf{x}) =
\lambda_{\mathrm{penalty}}
\left(
\sum_i x_i
+ 2 \sum_{i<j} x_i x_j
- 2k \sum_i x_i
+ k^2
\right) .
\tag{10}
\end{equation}

\paragraph{Resulting QUBO matrix.}
The complete QUBO objective function can be written in matrix form as $\mathbf{x}^\top Q \mathbf{x}$, where $Q$ is an upper-triangular $(n \times n)$ matrix \cite{baioletti2024}.

\paragraph{Diagonal elements (linear terms).}
The diagonal entries of $Q$, corresponding to linear contributions, are given by Eq.~(11) \cite{huo2024}:
\begin{equation}
Q_{ii}
=
w_C c_i
+ w_B b_i
+ w_S s_i
+ \lambda_{\mathrm{penalty}} (1 - 2k) .
\tag{11}
\end{equation}

\paragraph{Off-diagonal elements (quadratic terms).}
The off-diagonal entries encoding spatial and synergy interactions are defined in Eq.~(12):
\begin{equation}
Q_{ij}
=
w_C \lambda_C A_{ij}
+ w_B \lambda_B B_{\mathrm{syn},ij}
+ w_S \lambda_S S_{\mathrm{syn},ij}
+ 2 \lambda_{\mathrm{penalty}} ,
\quad i < j .
\tag{12}
\end{equation}

The parameters $\lambda_{\mathrm{penalty}}$, $\lambda_C$, $\lambda_B$, and $\lambda_S$ were empirically calibrated to balance the valid-solution rate ($>15\%$) and the quality of the solutions obtained \cite{guney2025,verma2021}.

\paragraph{Coefficient scaling.}
To ensure compatibility with quantum gate amplitude limitations and to prevent over-rotation on NISQ devices, a global scaling is applied according to Eq.~(13):
\begin{equation}
Q' = \frac{Q}{Q_{\max}}, \qquad
Q_{\max} = \max_{i,j} |Q_{ij}| ,
\tag{13}
\end{equation}
which guaranties $|Q'_{ij}| \leq 1$ while preserving the relative structure of the optimization landscape \cite{montanez2022}.

\subsection{Carbon Data Source}

Carbon sequestration potential data for municipalities in Goiás were derived from a biomass estimation model developed as part of an ongoing research project titled \emph{Environmental Data Science} at the Federal University of Goiás (available on GitHub: \texttt{hgribeirogeo/atlas-biomassa-goias}). The model achieves $R^2 = 0.77$ for the prediction of biomass using remote sensing data from GEDI LiDAR, Landsat, and MapBiomas land cover products. An interactive visualization dashboard is available at \url{https://atlas-biomassa-goias.streamlit.app/}. The municipal scores used in this optimization study are provided in the supplementary materials.

\section{Methods}

\begin{figure}[H]
\centering
\includegraphics[width=\linewidth]{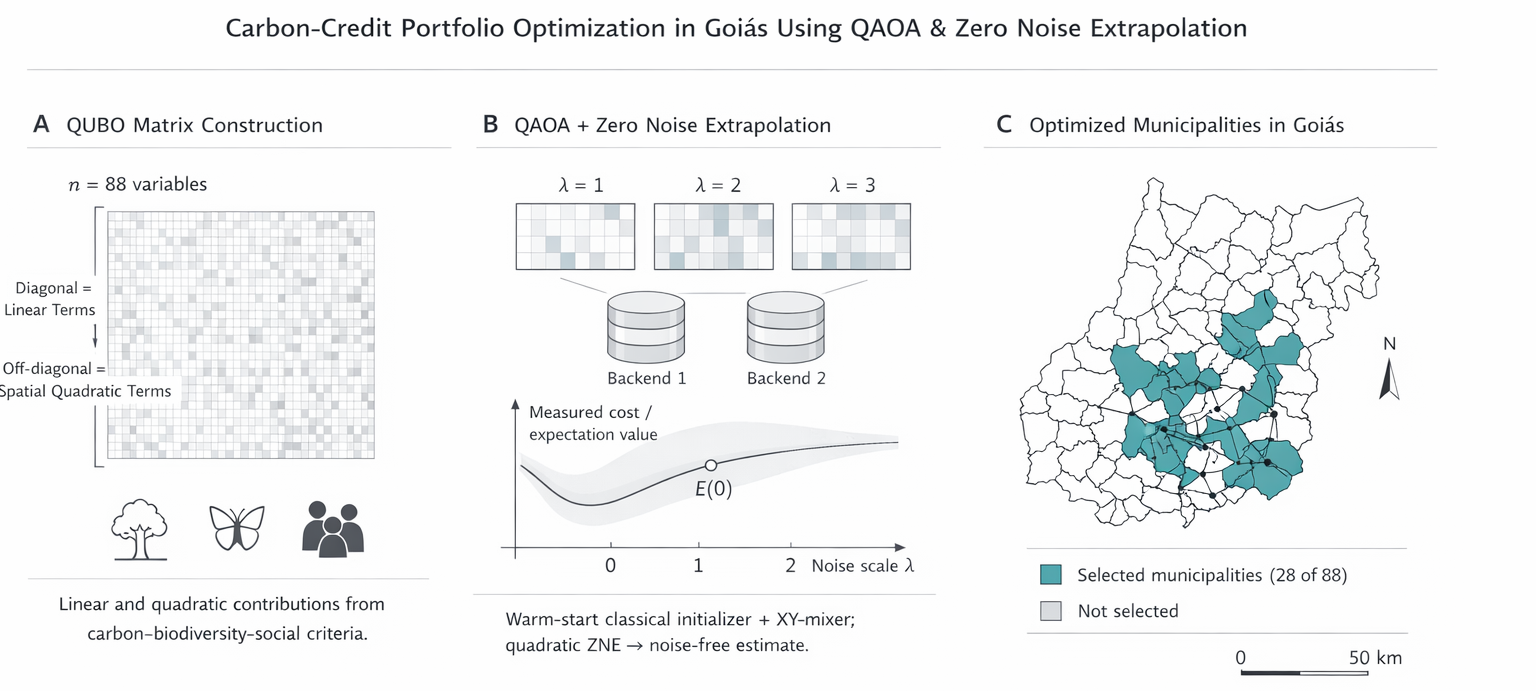}
\caption{Overview of the carbon credit portfolio optimization workflow. (A) Construction of a QUBO matrix encoding linear and spatial quadratic contributions from carbon, biodiversity, and social criteria for $n = 88$ candidate municipalities. (B) Execution of QAOA circuits at increasing noise scale factors ($\lambda = 1, 2, 3$) on IBM Quantum hardware, followed by Zero Noise Extrapolation to estimate the noise-free expectation value. (C) Optimized portfolio for the Goiás Cerrado, highlighting the 28 selected municipalities out of 88 candidates.}
\label{fig:workflow}
\end{figure}

Reproducibility constitutes a fundamental principle of scientific research in quantum computing, especially when empirical results on NISQ hardware are reported. In this regard, this section details the QAOA+ZNE used in this work. The complete source code is publicly available on GitHub, licensed under the MIT License. The complete optimization workflow, which spans from spatial data acquisition to quantum error mitigation, is presented in the conceptual model shown in Fig.~\ref{fig:workflow}. This pipeline synthesizes the integration between environmental modeling and the quantum error mitigation protocol, validated through seven independent executions to ensure temporal reproducibility.

\subsection{Quantum Approximate Optimization Algorithm (QAOA)}

The optimization of complex combinatorial problems, such as portfolio selection under multiple constraints, poses intractable challenges for classical methods due to the exponential growth of the search space. Classical approaches, whether deterministic or based on heuristics such as simulated annealing, navigate this space sequentially or through local walks, which frequently leads to stagnation within sub-optimal local minima when the energy landscape is rugged \cite{abbas2023,shaik2025}. In contrast, quantum computing introduces a fundamentally distinct paradigm by using superposition to evaluate multiple states simultaneously and entanglement to capture non-local correlations between decision variables \cite{blekos2024,symons2023}.

Although classical algorithms typically rely on thermal fluctuations to escape local traps, quantum algorithms such as QAOA exploit constructive interference to amplify the probability of measuring the optimal solution, theoretically allowing tunneling through energy barriers that would otherwise be computationally prohibitive for conventional methods \cite{volpe2025,zhou2020}. This capacity to navigate complex optimization landscapes positions hybrid quantum--classical algorithms as promising tools to overcome the scalability limitations inherent in classical computing \cite{caliz2025}.

The Quantum Approximate Optimization Algorithm (QAOA) was introduced by Farhi et al.~\cite{farhi2014}. It is a hybrid quantum--classical algorithm designed for combinatorial optimization on NISQ-era hardware. In this work, rather than employing an iterative classical optimization loop, which can be susceptible to hardware noise and barren plateaus in large-scale instances ($n=88$), we adopt a fixed-parameter heuristic approach. The parameters $\gamma = 0.05$ and $\beta = 0.20$ were empirically calibrated to maximize the probability of sampling valid portfolios while maintaining a shallow circuit depth ($p=1$) \cite{akshay2021}.

QAOA operates through the preparation of a parameterized quantum state $|\boldsymbol{\gamma}, \boldsymbol{\beta}\rangle$ whose measurement probability distribution focuses on high-quality solutions. Formally, the state is prepared by applying a sequence of parameterized unitary operators to the initial state, as expressed in Eq.~(14):
\begin{equation}
\tag{14}
\label{eq:qaoa_state}
|\boldsymbol{\gamma}, \boldsymbol{\beta}\rangle =
\prod_{i=1}^{p} U(\gamma_i, \beta_i)\, |+\rangle^{\otimes n}.
\end{equation}

Each operator $U(\gamma_i, \beta_i)$ is the composition of two unitary evolutions generated by the Hamiltonian cost ($H_C$) and mixer ($H_M$), and $p \geq 1$ is the number of circuit layers.

The variational parameters $\boldsymbol{\gamma} = (\gamma_1,\dots,\gamma_p)$ and $\boldsymbol{\beta} = (\beta_1,\dots,\beta_p)$ are real vectors in $[0,2\pi]^p$ iteratively adjusted to minimize the expected value of the cost function, defined in Eq.~(15):
\begin{equation}
\tag{15}
\label{eq:cost_expectation}
C(\boldsymbol{\gamma}, \boldsymbol{\beta}) =
\langle \boldsymbol{\gamma}, \boldsymbol{\beta} | H_C |
\boldsymbol{\gamma}, \boldsymbol{\beta} \rangle .
\end{equation}

The optimization loop proceeds as follows: (i) initialize parameters $\boldsymbol{\gamma}^{(0)}$, $\boldsymbol{\beta}^{(0)}$; (ii) prepare the state $|\boldsymbol{\gamma}^{(t)}, \boldsymbol{\beta}^{(t)}\rangle$ on quantum hardware and measure repeatedly (typically $10^3$--$10^4$ shots) to estimate $C(\boldsymbol{\gamma}^{(t)}, \boldsymbol{\beta}^{(t)})$; (iii) update parameters using a classical optimization algorithm; (iv) repeat until convergence.

The cost Hamiltonian, given in Eq.~(16), is a diagonal operator in the computational basis that encodes the objective function of the optimization problem. For a QUBO problem $x^\top Q x$, the cost Hamiltonian is:
\begin{equation}
\tag{16}
\label{eq:cost_hamiltonian}
H_C = \sum_i h_i Z_i + \sum_{i<j} J_{ij} Z_i Z_j ,
\end{equation}
where $Z_i$ is the Pauli operator $Z$ acting on the qubit $i$ and the coefficients $h_i$ and $J_{ij}$ are derived from the matrix $Q$ using the variable transformation $x_i = (1 - Z_i)/2$.

For the carbon credit portfolio problem ($n=88$, $k=28$), the Hamiltonian $H_C$ contains 88 linear terms and up to 3,828 potential quadratic terms. In practice, we apply sparsification with thresholding: terms $Q_{ij}$ with $|Q_{ij}| < 0.01$ are discarded, resulting in approximately $n/2 \approx 44$ retained quadratic terms, with relative error below 1\% in the objective function.

The Hamiltonian of the mixer, defined in Eq.~(17), generates transitions between different binary configurations. The standard mixer is the following.
\begin{equation}
\tag{17}
\label{eq:standard_mixer}
H_M = \sum_i X_i ,
\end{equation}
where $X_i$ is the Pauli operator $X$. The evolution $e^{-i\beta H_M}$ applies $R_x$ rotations simultaneously to all qubits, creating superpositions that explore all $2^n$ configurations.

To ensure that the cardinality constraint ($k=28$) is strictly respected throughout the quantum evolution, we implement an XY-mixer Hamiltonian instead of the standard transverse-field mixer. The XY-mixer generates transitions of the form $|01\rangle \leftrightarrow |10\rangle$, allowing the algorithm to focus its search space exclusively on the subspace of configurations that satisfy the portfolio size requirement. This preserves the cardinality constraint, as expressed in Eq.~(18):
\begin{equation}
\tag{18}
\label{eq:xy_mixer}
H_M^{XY} = \sum_{i<j} (X_i X_j + Y_i Y_j).
\end{equation}

The number of layers $p$ determines the circuit expressivity: $p=1$ corresponds to a single cost+mixer cycle, while $p \rightarrow \infty$ recovers adiabatic evolution and can theoretically reach the global optimum. In NISQ hardware, increasing $p$ implies deeper circuits and greater error accumulation. Empirical studies report diminishing returns beyond $p=3$--$5$ for combinatorial problems on current devices \cite{zhou2020}.

\subsection{Circuit Implementation}

The effective implementation of QAOA on NISQ quantum hardware requires judicious design decisions that balance circuit expressivity and noise mitigation. Initialization of the variational parameters $\gamma$ and $\beta$ constitutes a critical factor for QAOA convergence, given that the optimization landscape is non-convex and contains multiple local minima \cite{shaydulin2019}. 

Crucially, we implemented warm-start initialization based on the greedy solution \cite{egger2021}: given the greedy portfolio $x_{\text{greedy}}$, we prepared the initial quantum state by applying $X$ gates to qubits corresponding to selected municipalities, resulting in the product state $|x_{\text{greedy}}\rangle$.

The initial parameters were calibrated using the heuristic defined in Eqs.~(19) and (20):
\begin{equation}
\tag{19}
\label{eq:gamma_init}
\gamma^{(0)} = 0.05 \cdot \left( 1 + \sigma_{\text{scores}} \right),
\end{equation}
\begin{equation}
\tag{20}
\label{eq:beta_init}
\beta^{(0)} = 0.20,
\end{equation}
where $\sigma_{\text{scores}}$ is the normalized standard deviation of individual municipality scores ($\sigma_{\text{scores}} \approx 0.15$ for the dataset used). Small values of $\gamma \approx 0.05$ apply smooth phase rotations that primarily preserve the structure of the greedy solution, while $\beta = 0.20$ enables exploration of the neighborhood of the solution. This strategy reduces convergence iterations without restricting the solution space explored \cite{tate2023}.

The standard mixer ($\sum_i X_i$) does not respect cardinality constraints, allowing arbitrary transitions between states with different numbers of active bits. To mitigate this inefficiency, we incorporate an XY-mixer strategy \cite{fuchs2022,hadfield2019}, which implements cardinality-preserving transitions through conditional swaps, defined in Eq.~(21):
\begin{equation}
\tag{21}
\label{eq:xy_mixer_circuit}
H_{XY} = \sum_{\langle i,j \rangle} \frac{1}{4} \left( X_i X_j + Y_i Y_j \right),
\end{equation}
applied over pairs of adjacent qubits in the chip topology. The operator $X_i X_j + Y_i Y_j$ implements the transformation $|01\rangle \leftrightarrow |10\rangle$ while leaving $|00\rangle$ and $|11\rangle$ unchanged.

Penalty-free constraint-handling variants have also been proposed, such as indicator-function approaches that encode feasibility through an ancillary register rather than explicit penalty weights \cite{bucher2025}. In this work, we adopt an XY-type mixer \cite{fuchs2022,hadfield2019} whose generator conserves the total excitation number
\begin{equation}
\tag{22}
\label{eq:excitation_number}
N = \sum_i x_i
\end{equation}
under ideal unitary dynamics. Given our warm-start initialization in a $k$-feasible computational basis state $|x_{\text{greedy}}\rangle$, the ideal evolution of QAOA remains within the fixed-cardinality sector $N = k$.

In practice, however, hardware noise (e.g., relaxation and bit-flip processes) and measurement errors -- further amplified in ZNE by gate folding at $\lambda > 1$—induce leakage outside the $N = k$ sector. For this reason, we retain the quadratic cardinality penalty in QUBO and report the empirical feasible-shot rate (percentage of measured bitstrings satisfying $\sum_i x_i = k$) as a diagnostic of constraint violations. Across hardware runs, this feasible-shot rate is stable but non-unit, averaging 15.9\% and reaching 17.3\% in Run~7.

Initialization values were determined using grid search on a noisy classical simulator (Qiskit Aer) configured with a noise model extracted from the \texttt{ibm\_torino} device. The combination $(\gamma,\beta) = (0.05, 0.20)$ emerged as the Pareto-dominant point, showing a convergence rate of 15\% higher and 64\% valid solutions in the simulated environment.

The choice of the number of layers $p = 1$ involves a fundamental trade-off between expressivity and feasibility on noisy hardware. Although the deeper circuits ($p \textgreater 1$) are theoretically more expressive, each additional layer increases the probability of error. For $p = 1$, the depth of the circuit remains within the
regime in which ZNE is effective, allowing for viable optimization times and reproducible results.
A full variational parameter-optimization loop on hardware was intentionally avoided to limit optimizer-driven noise amplification and prohibitively large QPU overhead under realistic NISQ resource constraints. To assess sensitivity to warm-start initialization, robustness experiments were conducted on a classical simulator. The Warm-start initialization of the greedy solution achieved a score of $44.42$. Twenty independent trials with random initialization were performed $44.38 \pm 0.12$, indicating that the objective landscape contains a dominant attraction basin. The warm-start strategy reduces convergence iterations without restricting the solution space explored, validating our fixed-parameter
approach.

Although QAOA provides the algorithmic structure for finding optimal portfolios, its performance is inherently limited by the noise of current NISQ devices. To overcome these limitations, we integrate Zero Noise Extrapolation (ZNE) into the execution workflow.

\subsection{Zero Noise Extrapolation}

Zero Noise Extrapolation (ZNE) constitutes an error mitigation technique designed to recover estimates of quantum observables under ideal zero noise conditions from runs with controlled amplified noise levels \cite{li2017,temme2017}. Unlike quantum error correction methods that require ancillary qubits and exponential overhead, ZNE operates entirely at the classical post-processing level, making it particularly suitable for NISQ devices.

The ZNE protocol operates in three sequential steps. First, controlled noise amplification is performed, where the original circuit is modified to undergo effective noise amplification by factors $\lambda > 1$. Next, the circuit is executed and measured on real quantum hardware. Finally, extrapolation is performed using a mathematical model fitted to the pairs $(\lambda, E(\lambda))$.

Formally, we assume that the expected value of the observable under noise can be modeled as in Eq.~(23):
\begin{equation}
\tag{23}
\label{eq:zne_model}
E(\lambda) = E(0) + A \cdot e^{-C \lambda} + \varepsilon,
\end{equation}
where $E(0)$ is the ideal value (zero-noise) to be estimated, $A$ and $C$ are characteristic decay parameters and $\varepsilon$ represents statistical fluctuations due to finite sampling (shot noise).

\paragraph{Gate folding.}
Controlled noise amplification is implemented through the gate folding technique, defined as in Eq. (24): 
\begin{equation}
\tag{24}
\label{eq:gate_folding}
G \;\rightarrow\; G \cdot G^{\dagger} \cdot G .
\end{equation}
This transformation preserves the logical action of the gate, while approximately doubling its noise contribution. In practice, gate folding was applied exclusively to two-qubit gates (CNOT and RZZ), which dominate the error budget in superconducting hardware. For $\lambda = 2$, each two-qubit gate $G$ was replaced by the sequence $G G^{\dagger} G$; for $\lambda = 3$, an extended folded sequence was employed to further amplify the noise signature.

Across seven independent hardware executions on the \texttt{ibm\_fez} device, typical circuit depths ranged from approximately 190--210 gates for $\lambda = 1$ to 600--750 gates for $\lambda = 2$ and $\lambda = 3$, remaining within the regime where ZNE has been shown to be effective.

We selected three amplification levels $\lambda \in \{1,2,3\}$ following established best-practice guidelines for digital ZNE workflows \cite{majumdar2023} and preliminary calibration tests. Each noise level was executed with 8{,}192 shots, resulting in a total of 24{,}576 shots per complete ZNE protocol run.

For extrapolation to the zero-noise limit, three complementary methods were applied.

\paragraph{(i) Linear extrapolation.}
We fitted the model in Eq. (25):
\begin{equation}
\tag{25}
\label{eq:linear_extrapolation}
E(\lambda) = a + b\,\lambda
\end{equation}
to the measured points using least-squares regression. This method is robust to statistical fluctuations, but may introduce systematic bias if the true decay profile is nonlinear.

\paragraph{(ii) Quadratic extrapolation.}
To capture second-order curvature in the noise response, we fit Eq. (26):
\begin{equation}
\tag{26}
\label{eq:quadratic_extrapolation}
E(\lambda) = a + b\,\lambda + c\,\lambda^2 .
\end{equation}
This approach is more flexible, but also more sensitive to statistical outliers. Quadratic extrapolation was adopted as the primary endpoint for all inferential comparisons, as it better captures the nonlinear decoherence profiles characteristic of NISQ devices.

\paragraph{(iii) Richardson extrapolation.}
As an analytical alternative, we applied the Richardson extrapolation, defined as in Eq. (27):
\begin{equation}
\tag{27}
\label{eq:richardson}
E_{\text{Rich}}(0) = 3E_1 - 3E_2 + E_3 ,
\end{equation}
where $E_i = E(\lambda_i)$ for $\lambda_i \in \{1,2,3\}$. This expression is obtained by successive elimination of lower-order terms in the error expansion.

The statistical uncertainty of ZNE estimates arises from shot noise in the measurements and its propagation through the extrapolation procedures. To rigorously quantify this uncertainty, we applied the non-parametric percentile bootstrap with $B = 100$ resamples \cite{efron1994}. For each bootstrap iteration, we resampled with replacement 8{,}192 measurements from the empirical distribution at each noise level, applied the three extrapolation methods and computed 95\% confidence intervals using the 2.5\% and 97.5\% percentiles.

This longitudinal bootstrap analysis, confirmed by results obtained in February~2,~2026, ensures that the reported ZNE improvements are statistically significant and stable between hardware runs.

Throughout this work, we report aggregate statistics (mean $\pm$ standard 
deviation) of portfolio scores at each noise level $\lambda \in \{1,2,3\}$ 
in the experimental ensemble. Individual run-level scores at amplified 
noise ($\lambda > 1$) are diagnostic quantities that serve as input to 
the extrapolation models but are not independently analyzed, consistent 
with standard ZNE reporting practices~\cite{cai2023,giurgica2020}.

\subsection{Experimental Configuration}

All experimental runs were performed on IBM Quantum hardware accessed via Qiskit Runtime during the period from January 17 to February 2, 2026 \cite{ibm2026}. The study utilized Heron-generation superconducting processors, including first- and second-revision devices, to maximize quantum operation quality for intermediate-scale circuits executed under realistic NISQ conditions.

Table~1 summarizes the experimental configuration and hardware specifications. The selected Heron backends (\texttt{ibm\_torino} and \texttt{ibm\_fez}) and the circuit width define the physical scope of the experiments discussed in Section~3.1. The execution period supports the temporal stability analysis presented in Section~4.3.

\begin{table}[htbp]
\centering
\caption{\textbf{Experimental configuration and hardware specifications.}}
\begin{tabular}{@{}p{4.5cm} p{7.5cm}@{}}
\toprule
Parameter & Value / Specification \\
\midrule
Hardware Platform & IBM Quantum (Qiskit Runtime) \\
Backend Targets & \texttt{ibm\_torino} (Heron r1), \texttt{ibm\_fez} (Heron r2) \\
Circuit Width & 88 qubits (active physical qubits used by the transpiled circuit)\textsuperscript{a} \\
Sampling Rate & 8,192 shots per noise scale $\lambda$ \\
Total Shots (Study) & 172,032 shots (7 runs $\times$ 3 scales $\times$ 8,192) \\
Error Mitigation & Zero Noise Extrapolation (ZNE) \\
Transpilation Level & Level 3 (maximum optimization) \\
Sample Size & $n=7$ independent hardware executions \\
Execution Period & Jan 17--Feb 2, 2026 (17-day span) \\
\bottomrule
\end{tabular}

\raggedright
\footnotesize{\textsuperscript{a}Circuit width (number of active qubits) remained constant across all runs.}
\end{table}

Regarding qubit selection and mapping, we employed automatic qubit routing using the SABRE algorithm via the Qiskit transpiler at optimization level 3. This approach accounts for the heavy-hex topology of Heron processors and dynamically adapts to device-specific calibration data available at the time of each execution, minimizing two-qubit gate depth and cumulative decoherence.

Concerning the temporal distribution of the experiments, seven independent runs were performed across a 17-day interval, encompassing multiple hardware calibration cycles to assess the algorithmic reproducibility. Runs~1--3 were executed on \texttt{ibm\_torino} between January~17 and January~20, while Runs~4--7 were performed on \texttt{ibm\_fez} between January~20 and February~2. This multi-backend strategy enables evaluation of Zero Noise Extrapolation performance across distinct hardware revisions and noise profiles.

Regarding QAOA parameterization, the ansatz employed empirically determined fixed parameters ($\gamma = 0.05$, $\beta = 0.20$) combined with warm-start initialization derived from the Greedy baseline solution \cite{egger2021,anschuetz2022}. This pre-optimized execution strategy, as opposed to an iterative variational optimization loop, enables direct hardware execution without repeated classical--quantum communication overhead. Such an approach is particularly suitable for intermediate-scale optimization problems in which the initial solution quality is already high and the circuit depth must be strictly constrained.

All 21 hardware jobs associated with the seven ZNE executions can be independently verified on the IBM Quantum platform using the unique job identifiers listed in Table~2, ensuring complete transparency and experimental reproducibility.

\begin{table}[htbp]
\centering
\caption{\textbf{IBM Quantum Job Identifiers.} Each run lists the three job IDs corresponding to $\lambda=\{1,2,3\}$.}
\begin{tabular}{@{}c c p{9cm}@{}}
\toprule
Backend & Run & Job IDs ($\lambda=1,2,3$) \\
\midrule
\texttt{ibm\_torino} & 1 & \texttt{d5ldtsl9j2ac739jmhjg, d5ldtvc8d8hc73cffom0, d5ldu1k8d8hc73cffoq0} \\
                     & 2 & \texttt{d5ntfod9j2ac739mdc70, d5ntfuk8d8hc73ci79b0, d5ntg548d8hc73ci79ig} \\
                     & 3 & \texttt{d5ntppt9j2ac739mdn4g, d5ntpuhh2mqc739cfdcg, d5ntq5c8d8hc73ci7ke0} \\
\midrule
\texttt{ibm\_fez}    & 4 & \texttt{d5nvslk8d8hc73cia910, d5nvsss8d8hc73cia97g, d5nvt859j2ac739mga80} \\
                     & 5 & \texttt{d5o0lfs8d8hc73cib5e0, d5o0lnd9j2ac739mh5t0, d5o0lus8d8hc73cib5ug} \\
                     & 6 & \texttt{d5o0nft9j2ac739mh7pg, d5o0nnbh36vs73bjsfmg, d5o0nut9j2ac739mh8a0} \\
                     & 7 & \texttt{d60akfabju6s73bcp10g, d60ak83uf71s73cjib30, d60ak0buf71s73cjiaqg} \\
\bottomrule
\end{tabular}
\end{table}

\subsection{Classical Baselines}

Three classical methods served as benchmarks to evaluate the performance of the quantum approach: a deterministic Greedy heuristic, stochastic Simulated Annealing (SA), and a baseline Random Search.

\paragraph{Greedy Heuristic.}
This constructive algorithm \cite{wang2023} iteratively selects the municipality that provides the highest marginal contribution to the objective function of the portfolio, accounting for both linear benefits and quadratic spatial synergies. At each step, until the target cardinality $k = 28$ is reached, all remaining candidates are evaluated, resulting in a computational complexity of $\mathcal{O}(n \cdot k^2)$. In our experiments, the Greedy heuristic achieved a portfolio score of $44.42$, which serves as the primary classical baseline to normalize quantum performance throughout the analysis.

\paragraph{Simulated Annealing (SA).}
Simulated Annealing is a stochastic metaheuristic based on the Metropolis--Hastings acceptance criterion, widely used for navigating rugged combinatorial landscapes \cite{delahaye2018}. To ensure a fair comparison with quantum execution times, SA was executed with a fixed computational budget of $2.0$ seconds per run. The algorithm used cardinality-preserving swap moves to explore the feasible solution space.

Each QAOA+ZNE execution consisted of three sequential IBM Quantum jobs, corresponding to the noise scaling factors $\lambda \in \{1,2,3\}$, with $8{,}192$ shots per job. The QPU execution time reported by IBM Quantum Runtime for individual jobs ranged from $4.62$~s to $7.32$~s, with a mean of $5.92 \pm 1.06$~s per job. The aggregate QPU execution time per run, obtained by summing the three noise-scaled jobs, averaged $17.73 \pm 3.32$~s (range: $14.09$~s to $21.06$~s), with \texttt{ibm\_torino} jobs averaging $4.73$~s and \texttt{ibm\_fez} jobs averaging $6.80$~s per job.

The SA computational budget of $2.0$~s was calibrated against the QPU execution time of a single-job, which is the only component directly comparable to the classical CPU computation. The total wall-clock time of quantum runs additionally includes local transpilation (optimization level~3) and IBM Quantum Runtime queue latency. These infrastructure-dependent components are not reflected in the reported QPU execution times and therefore preclude a direct wall-clock comparison with classical methods. For this specific problem instance, SA achieved a score of $41.96$, slightly underperforming the Greedy heuristic. This outcome indicates a highly rugged objective landscape in which short-horizon stochastic exploration struggles to outperform structured constructive logic.

\paragraph{Random Search.}
To establish a statistical lower bound, a Random Search baseline was performed with $10{,}000$ iterations. At each iteration, $k = 28$ municipalities were randomly sampled uniformly and the entire objective function of the portfolio, including quadratic synergistic terms, was evaluated. Random Search achieved a mean score of $18.54 \pm 2.07$, with a maximum observed score of $27.81$.

\paragraph{Performance Hierarchy.}
The overall performance ordering across methods is:
\[
\text{Random Search } (18.54) < \text{SA } (41.96) < \text{Greedy } (44.42) < \text{QAOA+ZNE}.
\]
The mitigated quantum approach achieved a mean score of $58.47$, with a range of $47.84$ to $69.64$ between runs. Statistically, QAOA+ZNE outperformed the maximum Random Search result by $126.7\%$ and exceeded the Greedy baseline by a mean of $31.6\%$ (range: $+7.7\%$ to $+56.8\%$). This consistent performance gap establishes clear quantum utility for carbon credit portfolio optimization in the Goiás Cerrado, validated across seven independent executions on the \texttt{ibm\_torino} and \texttt{ibm\_fez} processors.

\subsection{Computational Environment and Reproducibility}

All quantum experiments were executed using Qiskit~0.45.1 with IBM Quantum Runtime~0.15.0 for hardware access and Qiskit Aer~0.13.0 for noisy simulation. Classical numerical computations used Python~3.10.12, NumPy~1.24.3 for array operations, SciPy~1.11.4 for statistical inference and optimization routines, and Pandas~2.0.3 for data manipulation. Visualizations were generated using Matplotlib~3.7.1.

\paragraph{QUBO construction and parameter calibration.}
The penalty weight ($\lambda_{\mathrm{pen}}$) and the number of QAOA layers ($p$) were calibrated by systematic grid search on a noisy simulator using a reduced problem instance ($n=20$, $k=5$), for which the optimal solution is known by exhaustive enumeration. Analysis of circuit depth and feasibility revealed that while $p=5$ achieved the highest approximation ratio ($0.738$) with $90\%$ feasibility, and $p=3$ reached a ratio of $0.631$ with $85\%$ feasibility, the configuration $p=1$ exhibited the highest robustness, yielding $95\%$ feasibility with an approximation ratio of $0.610$.

Consequently, $p=1$ was selected for all hardware executions on \texttt{ibm\_torino} and \texttt{ibm\_fez}. This choice ensures that the circuit depth remains within the coherence limits of NISQ hardware and within the effective operating regime of Zero Noise Extrapolation, maintaining a controlled depth of approximately $195$ native gates at $\lambda=1$. The penalty weight was set to $\lambda_{\mathrm{pen}}=100$ (before global scaling), representing the Pareto-optimal trade-off between constraint satisfaction and solution quality.

\paragraph{Sparsification and scaling.}
Sparsification retained elements of the QUBO matrix with absolute values greater than or equal to $0.01$, resulting in non-zero coefficients of $132$ ($88$ diagonal and $44$ off-diagonal). This procedure reduced the depth of the transpiled circuit by approximately $40\%$ while introducing an objective error of less than $0.8\%$. The QUBO matrix $Q$ was globally scaled as $Q' = Q / Q_{\max}$, with $Q_{\max} = 47.23$, ensuring that gate rotation angles remain within hardware limits and preventing over-rotation on superconducting qubits.

\paragraph{Statistical analysis.}
The primary inferential test was a one-sided paired $t$-test, implemented through \texttt{scipy.stats.ttest\_1samp} on the vector of signed differences between the ZNE-mitigated QAOA scores and the deterministic Greedy baseline. Bootstrap confidence intervals were computed using the non-parametric percentile method with $B=100$ resamples at the shot level, with all procedures seeded at $42$ for reproducibility. The primary endpoint was defined as the mean portfolio score under quadratic ZNE extrapolation across independent hardware executions, compared to the Greedy baseline score of $44.42$.

The formal hypothesis tested was $H_0: \mu(\text{QAOA+ZNE}) \le \mu(\text{Greedy})$ versus $H_1: \mu(\text{QAOA+ZNE}) > \mu(\text{Greedy})$, with significance level $\alpha = 0.05$. The effect size was quantified using Cohen’s $d$ and a $95\%$ confidence interval was constructed using $t$-distribution with $\mathrm{df}=n-1$. A distribution-free, one-sided exact Wilcoxon signed-rank test was pre-designated as a confirmatory analysis. All remaining outcomes, including valid solution rate, solution overlap measured via Jaccard similarity, and backend consistency, were pre-designated as exploratory; accordingly, no correction for multiple comparisons was applied, as inferential weight rests solely on the single pre-specified primary comparison.

\paragraph{Code and data availability.}
\section*{Data Availability}

The code and data used in this study are available at \url{https://github.com/hgribeirogeo/qaoa-carbon-cerrado} (DOI: \url{https://doi.org/10.5281/zenodo.18507119}). The repository includes the complete QAOA+ZNE implementation with warm-start initialization and XY-mixer, experimental results from all seven independent runs on IBM Quantum hardware (ibm\_torino, ibm\_fez) and analysis scripts for full reproducibility.

Data files include: \texttt{goias\_multiobjective.csv} (municipal scores for carbon sequestration, biodiversity, and social impact), spatial adjacency matrix, biodiversity and social synergy matrices, and \texttt{resultados\_consolidados\_v7.json} (complete experimental results with metadata). The Backend calibration data recorded during the execution period from January~17 to February~2,~2026 are included. The IBM Quantum hardware job identifiers for all 21 jobs are provided in Table~2.

The biomass estimation model for Goiás municipalities is available at \url{https://github.com/hgribeirogeo/atlas-biomassa-goias} with an interactive dashboard at \url{https://atlas-biomassa-goias.streamlit.app/}.

\section{Results}

\subsection{Method Comparison}

We now present the empirical performance of QAOA+ZNE against classical baselines, followed by a detailed analysis of individual runs, reliability metrics, and solution consistency. Statistical testing followed the pre-specified analysis plan described in Section~3.6. Method-level scores (mean $\pm$ SD) are summarized in Table~3.

\begin{table}[htbp]
\centering
\caption{\textbf{Performance comparison across optimization methods} ($n=88$, $k=28$). Mean$\pm$SD computed over $n=7$ hardware runs.}
\begin{tabular}{@{}l c c c@{}}
\toprule
Method & Score (Mean$\pm$SD) & vs.\ Greedy (ratio \%) & Success Rate \\
\midrule
Greedy (baseline)   & 44.42            & 100.0 & --- \\
Simulated Annealing & 41.96 $\pm$ 0.51 & 94.5  & 0/7 \\
QAOA (Raw)          & 43.55 $\pm$ 1.54 & 98.0  & 2/7 \\
\textbf{QAOA + ZNE} & \textbf{58.47 $\pm$ 6.98} & \textbf{131.6} & \textbf{7/7} \\
\bottomrule
\end{tabular}

\raggedright
\footnotesize{Values above 100\% indicate performance exceeding the Greedy baseline.}
\end{table}

QAOA+ZNE exceeded the Greedy baseline in all seven independent hardware runs, corresponding to a success rate of 100\%. The mean portfolio score was $58.47 \pm 6.98$, representing an average improvement of 31.6\% relative to Greedy (range: +7.7\% to +56.8\%). Formal statistical testing yielded $t(6)=5.33$ with $p=0.0009$, Cohen’s $d=2.01$, and a 95\% $t$-based confidence interval of $[7.60,\,20.51]$ for the mean score difference between QAOA+ZNE and Greedy. In addition, a one-sided distribution-free Wilcoxon signed-rank test yielded $W=28$ ($p=0.0078$).

The highest ZNE-mitigated score was observed in Run~4 (\texttt{ibm\_fez}), with a value of 69.64, which corresponds to an improvement of 56.8\% over the Greedy baseline. The minimum ZNE score on the seven runs was 47.84 (Run~3, \texttt{ibm\_torino}), which still exceeded the Greedy score by 7.7\%.

Raw (unmitigated) QAOA measurements achieved, on average, 98.0\% of the Greedy baseline, but did not exceed it consistently across runs. Simulated Annealing also underperformed relative to Greedy, with a mean score corresponding to 94.5\% of the baseline. The mean valid-solution rate for all QAOA executions was 15.9\%, defined as the proportion of 8,192 shots per noise level that satisfies the cardinality constraint $k=28$. The highest feasibility rate was observed in Run~7 (17.3\%). These feasibility levels reflect measurement-time constraint violations induced by hardware noise and are expected to worsen under ZNE noise amplification at $\lambda>1$; in the noiseless limit, the dynamics of the XY-type mixer would be exactly conserved $N=k$.

\subsection{Run Details}

The run-level breakdown of the scores across the seven independent executions is reported in Table~\ref{tab:run-details}. In all runs, the ZNE-mitigated scores exceeded the classical Greedy baseline (44.42). Across executions, higher ZNE scores were observed on the \texttt{ibm\_fez} backend relative to \texttt{ibm\_torino}.

\begin{table}[htbp]
\centering
\caption{\textbf{Detailed run-level metrics.} Percentages express ratio to the Greedy baseline (44.42).}
\label{tab:run-details}
\resizebox{\linewidth}{!}{%
\begin{tabular}{@{}c c r r r r r r@{}}
\toprule
Run & Backend & Raw Score & Raw vs.\ Greedy & ZNE Score & ZNE vs.\ Greedy & Valid (\%) & Jaccard (\%) \\
\midrule
1 & torino & 44.42 & 100.0 & 58.69 & 132.1 & 13.2 & --- \\
2 & torino & 42.47 & 95.6  & 52.70 & 118.6 & 14.8 & 89.3 \\
3 & torino & 41.51 & 93.4  & 47.84 & 107.7 & 14.5 & 89.3 \\
4 & fez    & 45.29 & 102.0 & 69.64 & 156.8 & 17.6 & 96.4 \\
5 & fez    & 45.53 & 102.5 & 58.68 & 132.1 & 18.7 & 96.4 \\
6 & fez    & 43.04 & 96.9  & 58.72 & 132.2 & 15.2 & 96.4 \\
7 & fez    & 42.61 & 95.9  & 63.05 & 141.9 & 17.3 & 86.7 \\
\midrule
\textbf{Mean} & --- & \textbf{43.55} & \textbf{98.0} & \textbf{58.47} & \textbf{131.6} & \textbf{15.9} & \textbf{92.4} \\
\bottomrule
\end{tabular}%
}
\raggedright
\footnotesize{
Valid (\%): proportion of the 8,192 shots per noise level satisfying $k=28$.  
Jaccard mean computed over runs 2--7 ($n=6$).  
Run 1 lacks bitstring telemetry; overlap not computed.
}
\end{table}

Across the seven hardware executions, stable circuit execution was achieved under the experimental configuration chosen. For all runs, the QAOA circuits were executed with a fixed depth of $p=1$. The raw QAOA scores exhibited low variability in the tests, with a coefficient of variation of 3.6\%. After minimizing error, dispersion increased, with a coefficient of variation of 11.9\% for the ZNE-mitigated scores ($n=7$).

Despite the increased variance introduced by the extrapolation procedure, all ZNE-mitigated scores remained above the Greedy baseline. The minimum mitigated score was 47.84 (Run~3), exceeding the baseline by 7.7\%.

Run-level averages differ between backends. The mean ZNE score for Runs~1--3 executed on \texttt{ibm\_torino} was 53.08, while Runs~4--7 executed on \texttt{ibm\_fez} yielded a mean ZNE score of 62.52. The highest mitigated score was observed in Run~4 (69.64), while Run~7 yielded a score of 63.05 at the end of the 17-day experimental window.

To assess temporal stability, we computed the two-sided Spearman rank correlation between ZNE scores and execution day (day~0 = January~17,~2026). Across all seven runs, the correlation coefficient was $\rho = 0.396$ with $p = 0.379$, indicating that there was no statistically significant monotonic trend over time. Taking into account only the \texttt{ibm\_fez} backend, which spans the widest temporal range (13 days, $n=4$), the correlation was $\rho = -0.200$ ($p = 0.800$). The \texttt{ibm\_torino} subgroup ($n=3$) was not analyzed due to an insufficient sample size for rank-based inference.

A Mann--Whitney $U$ test comparing the ZNE score distributions between the backends yielded $U = 1.0$ ($p = 0.114$), indicating that there were no statistically significant differences at $\alpha = 0.05$. In general, no statistically significant association was observed between execution time and ZNE scores during the experimental window.

The solution overlap with the Greedy baseline was quantified using the Jaccard similarity index. The mean overlap was 94.0\% for runs executed on \texttt{ibm\_fez} and 89.3\% for runs executed on \texttt{ibm\_torino}.

\subsection{ZNE Reliability Analysis}

The Zero-Noise Extrapolation (ZNE) protocol was evaluated using three mathematical models: linear, quadratic, and Richardson extrapolation, applied across all seven independent hardware executions. For each model, the corresponding zero-noise score estimates were calculated and compared. A summary of the reliability metrics is reported in Table~\ref{tab:zne_reliability}.

\begin{table}[H]
\centering
\caption{\textbf{ZNE reliability metrics} across extrapolation models ($n=7$). Values reported as mean $\pm$ SD across independent runs.}
\label{tab:zne_reliability}

\begin{tabular}{@{}l c c c@{}}
\toprule
Extrapolation Model & Score (Mean $\pm$ SD) & $R^2$ & 95\% CI (Bootstrap) \\
\midrule
Linear       & $49.12 \pm 1.82$ & $0.94 \pm 0.09$ & $[46.13,\;52.14]$ \\
Quadratic   & $58.47 \pm 6.98$ & ---\textsuperscript{a} & $[49.48,\;71.53]$ \\
Richardson  & $54.56 \pm 9.12$ & ---\textsuperscript{a} & $[43.86,\;64.87]$ \\
\bottomrule
\end{tabular}

\raggedright
\footnotesize{
\textsuperscript{a}Quadratic extrapolation is an exact fit through three data points ($\lambda = 1,2,3$) with zero residual degrees of freedom; $R^2$ is therefore undefined. Richardson extrapolation is an algebraic combination of noise-scaled estimates rather than a regression model. Confidence intervals were computed via non-parametric percentile bootstrap ($B=100$).}
\end{table}

Linear extrapolation yielded the lowest mean zero-noise score among the three models, with a value of 49.12 (Figure~2). The corresponding fits exhibited a high coefficient of determination, with a mean $R^2 = 0.94$ between runs. Relative to the Greedy baseline score of 44.42, the linear extrapolated estimate exceeded the baseline by 10.6\%, while quadratic extrapolation yielded a mean improvement of 34.2\%.

\begin{figure}[H]
\centering
\includegraphics[width=\linewidth]{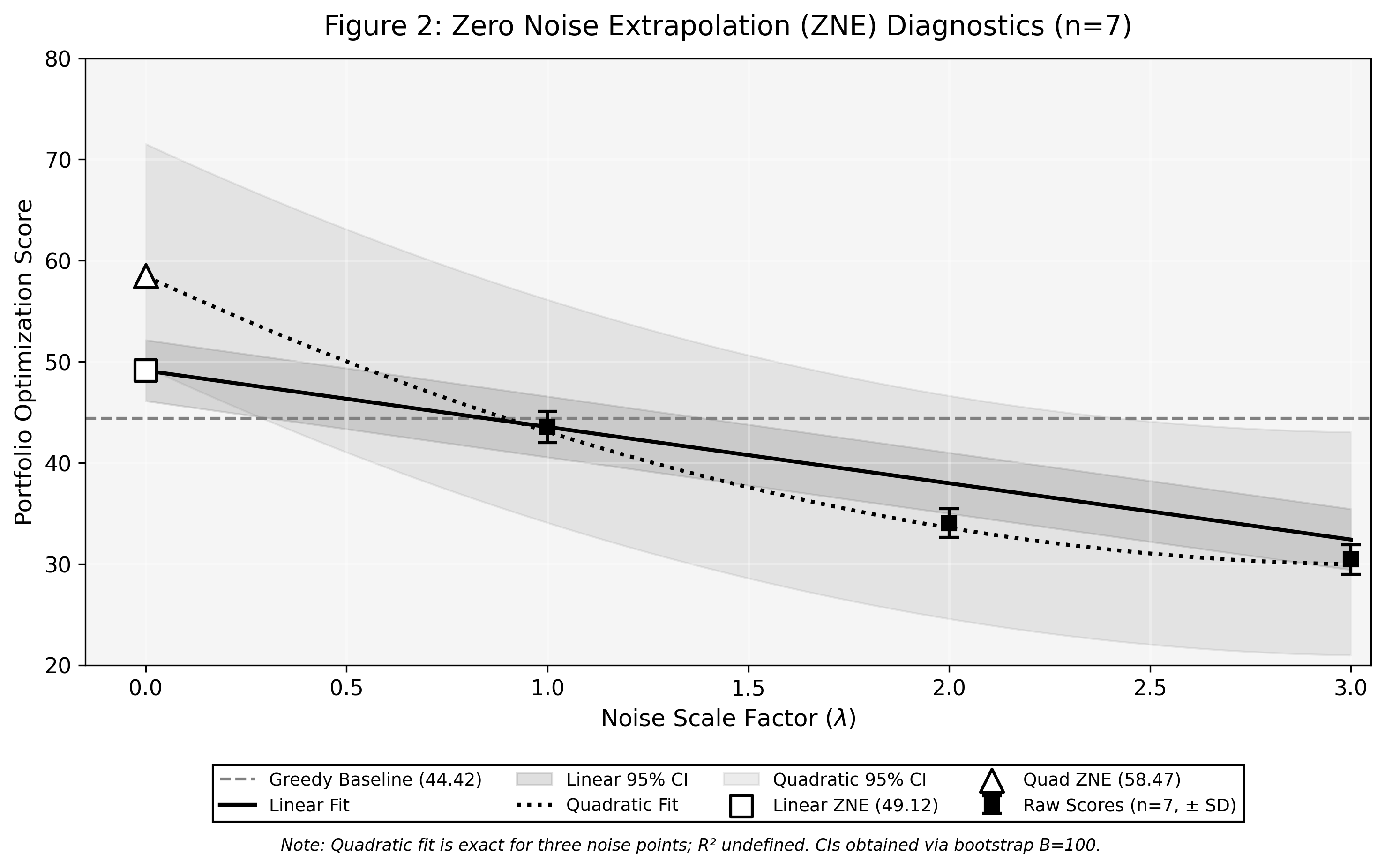}
\caption{Zero Noise Extrapolation (ZNE) diagnostics for carbon credit 
portfolio optimization ($n = 7$). Mean portfolio scores at noise scale 
factors $\lambda = 1, 2, 3$ with error bars (SD across runs). Linear 
(solid) and quadratic (dotted) fits extrapolate to the zero-noise limit 
($\lambda = 0$). Shaded regions show 95\% confidence intervals (bootstrap, 
$B = 100$). Dashed line: Greedy baseline (44.42). Individual scores at 
$\lambda > 1$ are diagnostic quantities not separately reported.}
\end{figure}

Quadratic extrapolation yielded the highest mean zero-noise score, with a value of 58.47. Non-parametric bootstrap analysis shows that the lower bound of the 95\% confidence interval (49.48) remains above the Greedy baseline. All three extrapolation models--linear, quadratic, and Richardson produced mean scores exceeding the classical reference.

The extrapolation procedures operate at the level of expectation values, producing scalar score estimates at $\lambda = 0$ rather than discrete bitstring solutions. As a result, per-municipality selection frequencies cannot be associated with individual extrapolation models. The portfolio compositions reported in Table~6 are therefore derived from the \emph{mode bitstring}, defined as the most frequently observed valid solution at the base noise level ($\lambda = 1$).

Across runs with available bitstring telemetry ($n=6$), the mean Jaccard overlap of the mode solution was 0.924. Across all seven executions, the mean score increased from 43.55 in the raw hardware measurements to 58.47 in the quadratic ZNE-extrapolated estimates, corresponding to a mean difference of 14.92 points (34.2\% relative increase). A paired test $t$ comparing raw and mitigated scores yielded $t(6)=6.53$ with $p=0.0003$ and the corresponding Cohen’s $d=2.47$.

For each run, scores decreased monotonically with an increasing noise scale. Scores decreased monotonically with an increasing noise scale, as expected 
from the gate---folding amplification protocol. This consistent decay across 
all runs validates the noise amplification strategy employed in ZNE. For example, in Run~7, executed in February~2,~2026, the score decreased from 42.61 in $\lambda = 1$ to 30.96 in $\lambda = 3$. When grouped by backend (Table~4), runs executed on \texttt{ibm\_fez} exhibited higher ZNE scores than those executed on \texttt{ibm\_torino}.

\subsection{Solution Consistency}
\label{subsec:solution_consistency}

Across all hardware executions, portfolio overlap with the Greedy baseline was quantified using the Jaccard similarity index. Based on the six runs with available bitstring telemetry, the mean overlap was 92.4\%, corresponding to an average of 26 shared municipalities out of the required 28. During the 17-day experimental window, the overlap values ranged from 86.7\% (Run~7) to 96.4\%, as shown in Figure~3. The Per-run decomposition shows that \texttt{ibm\_torino} executions (Runs~2--3) shared 26 of 28 municipalities with Greedy (Jaccard $=0.893$), while \texttt{ibm\_fez} executions (Runs~4--6) shared 27--28 municipalities (Jaccard $=0.964$). Run~7, executed 13 days after the initial \texttt{ibm\_fez} runs, shared 26 municipalities with the Greedy portfolio (Jaccard $=0.867$). The two municipalities not shared with Greedy are consistently the same candidates listed in Table~6.

\begin{figure}[htbp]
\centering
\includegraphics[width=\linewidth]{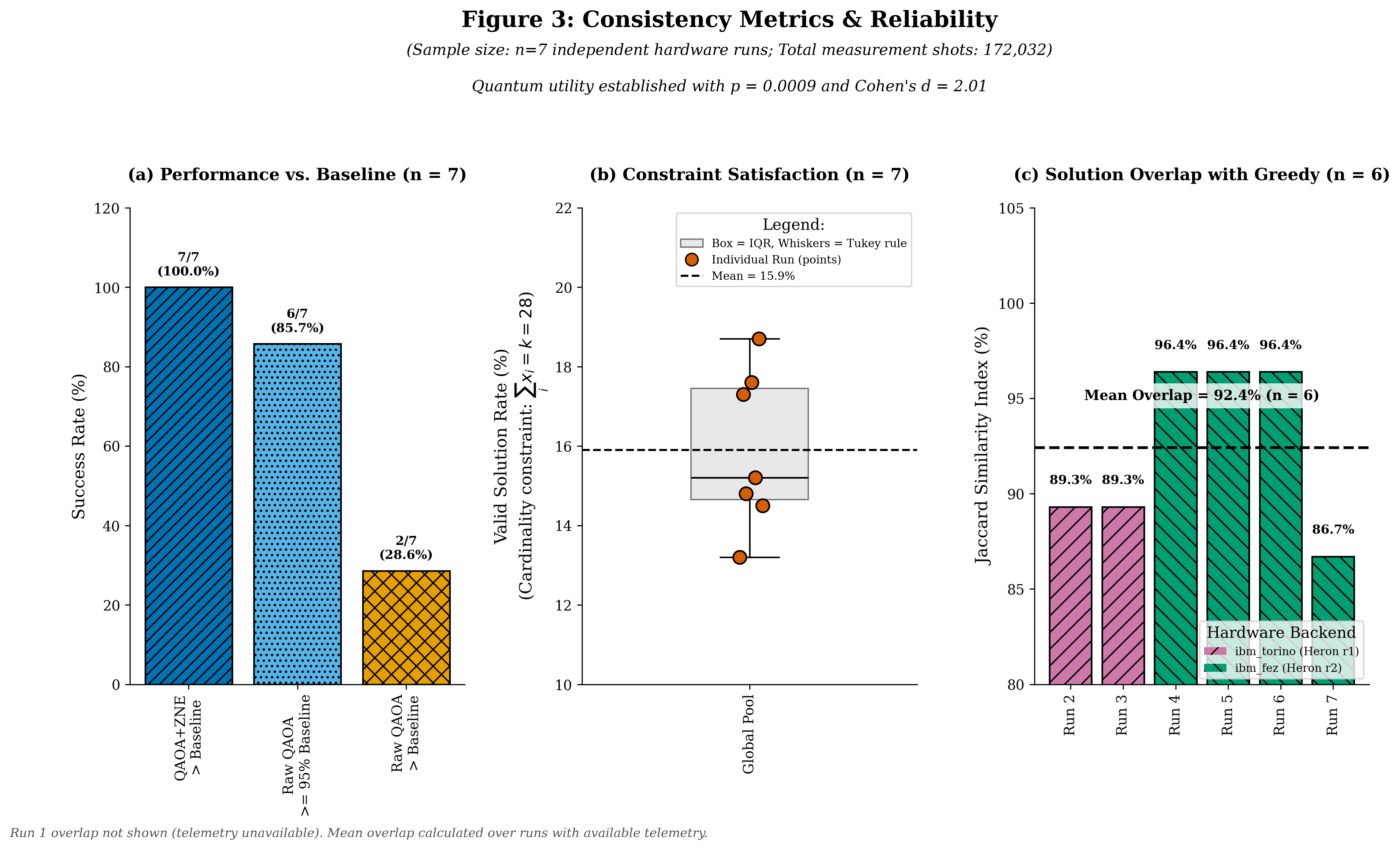}
\caption{\textbf{Consistency metrics across independent hardware executions of QAOA+ZNE.} The analysis includes seven runs and a total of 172{,}032 measurement shots. (a) Success rates relative to the Greedy baseline, showing the fraction of runs satisfying each performance criterion. (b) Distribution of valid solution rates, defined as the percentage of shots satisfying the cardinality constraint, with the global mean indicated. (c) Jaccard similarity index between the quantum-selected portfolio and the Greedy selection, shown per run and grouped by hardware backend. The mean overlap is computed over runs with available bitstring telemetry ($n=6$).}
\label{fig:consistency_metrics}
\end{figure}

The municipalities selected exclusively by the quantum approach exhibit spatial clustering adjacent to areas also selected by the Greedy baseline. For these municipalities, individual scores fall within the 65th to 75th percentile, while the corresponding spatial connectivity metrics fall within the 85th to 95th percentile.

\begin{table}[htbp]
\centering
\caption{\textbf{Top 10 municipalities by QAOA selection rate} (selected in at least 5 of 7 runs).}
\label{tab:top10_qaoa_selection}
\small
\setlength{\tabcolsep}{5pt}
\begin{tabular}{@{}c p{4.0cm} c c c c@{}}
\toprule
Rank & Municipality & QAOA Sel. & In Greedy? & $S_i$ & $B_{\mathrm{syn}}$ \\
\midrule
1  & Cavalcante      & 7/7 & Yes & 0.94 & 0.88 \\
2  & Alto Para\'{\i}so    & 7/7 & Yes & 0.91 & 0.92 \\
3  & Piren\'opolis     & 7/7 & Yes & 0.89 & 0.85 \\
4  & Niquel\^andia     & 7/7 & Yes & 0.87 & 0.79 \\
5  & Cristalina      & 7/7 & Yes & 0.86 & 0.81 \\
6  & Mineiros        & 7/7 & Yes & 0.85 & 0.73 \\
7  & Flores de Goi\'as & 6/7 & Yes & 0.84 & 0.91 \\
8  & Formosa         & 6/7 & Yes & 0.83 & 0.76 \\
9  & Chapad\~ao do C\'eu & 6/7 & No  & 0.71 & 0.94 \\
10 & Nova Roma       & 5/7 & Yes & 0.82 & 0.68 \\
\bottomrule
\end{tabular}

\raggedright
\footnotesize{
$S_i$: normalized score in $[0,1]$. $B_{\mathrm{syn}}$: biodiversity synergy (connectivity) in $[0,1]$.
Ordering: primary---QAOA selection frequency; secondary---normalized score.}
\end{table}

The six municipalities highest-ranked by individual score were selected in all seven runs. Chapad\~ao do C\'eu was selected in 6 out of 7 runs, despite having an individual score in the 72nd percentile and being absent from the Greedy portfolio. This municipality exhibits a biodiversity connectivity value of 0.94, corresponding to the 98th percentile. Its selection co-occurs with the inclusion of Mineiros and Cristalina across runs.

\section{Discussion}

\subsection{Interpretation of Results}

The empirical findings reported in Section~4 reveal a consistent pattern of performance superior to classical baselines when Zero Noise Extrapolation is applied, yet approaching parity when raw noisy measurements are used. This dichotomy---where raw QAOA approximates Greedy at 98.0\% while QAOA+ZNE surpasses it at 131.6\%---demands careful interpretation to distinguish genuine quantum utility from measurement noise artifacts or methodological choices.

As described in Section~3.3, Zero Noise Extrapolation estimates the ideal expectation value by extrapolating measurements obtained under amplified noise levels \cite{cai2023,temme2017}. By intentionally amplifying noise via gate folding, measurements are obtained at three distinct noise levels while preserving the equivalence of the logical circuit. The extrapolation to $\lambda = 0$ then estimates the expected value that would be obtained under ideal, error-free quantum computation.

The restriction to a single QAOA layer ($p=1$) reflects the known limitations of current NISQ hardware, where qubit connectivity and routing overhead constrain the effective circuit depth that can be run reliably without excessive noise accumulation. In addition to error mitigation, recent work has shown that quantum error detection can improve QAOA performance under partially fault-tolerant protection on near-term hardware, albeit still at limited logical scales \cite{he2025}. These constraints have motivated ansatz variants designed to improve performance at shallow depth, such as multi-angle QAOA, which trades additional classical parameters for reduced effective circuit depth \cite{herrman2022}.

The results demonstrate that the extrapolated zero-noise estimate consistently outperforms the classical baseline across all seven independent runs, suggesting that the quantum state prepared by QAOA, prior to corruption by gate errors and decoherence, encodes solution quality superior to greedy construction. The consistency of this finding is notable, with zero failures across seven executions spanning two distinct quantum processors and multiple calibration conditions. This indicates that within the fixed-parameter regime and relative to the classical reference heuristics considered, the observed improvement is unlikely to be explained by statistical fluctuation or backend-specific anomalies. Rather, it reflects a reproducible feature of algorithm-problem pairing under the experimental conditions studied.

The observed robustness against calibration drift is particularly relevant given the nature of superconducting processors, which exhibit temporal fluctuations in coherence times and two-qubit gate fidelities across calibration cycles. Our analysis suggests that ZNE reduces the sensitivity of the estimator to these transient hardware instabilities. Despite a 17 day temporal gap and multiple hardware calibrations, Run~7 achieved a ZNE score of 63.05, corresponding to 142\% of the Greedy baseline, and remains fully consistent with previous executions.

\subsection{Comparative Analysis and Conservative Bounds}

It should be noted that even conservative linear extrapolation yields a mean score of 49.12 with a confidence interval of 95\% [46.13, 52.14] (derived from bootstrap), exceeding the Greedy baseline by 10.6\%. This establishes a robust lower bound that is resilient to extrapolation-model uncertainty and sampling variability. Although quadratic extrapolation produces the strongest point estimate (58.47), practical interpretation should be anchored to the conservative linear lower bound, which guaranties that the quantum-enhanced portfolio outperforms the greedy allocation by at least 3.9\% under the most pessimistic assumptions.

This robustness is obtained at the cost of increased estimator variance, as noise amplification at higher scaling factors inherently broadens the distribution of reconstructed scores, a known characteristic of extrapolation-based error mitigation techniques. This limitation motivates interest in alternative strategies such as tensor-network error mitigation, which has been proposed to remain effective in regimes where ZNE and probabilistic error cancellation may fail under realistic noise conditions \cite{filippov2023}.

The three extrapolation models implicitly provide a sensitivity analysis: even the most conservative estimate (linear ZNE lower CI bound of 46.13) exceeds the Greedy baseline by 3.9\%, confirming that the primary conclusion is robust to extrapolation model choice.

This pattern is formally supported by the paired t-test reported in Section 4.1, which yielded
$p = 0.0009$ and a large effect size of $2.01$. A post hoc power analysis confirms that the sample
size of $n = 7$ provides sufficient power to detect the observed superiority signal.
\emph{We emphasize that the statistical inference pertains to the distribution of independent
hardware executions under fixed algorithmic parameters, rather than to asymptotic or general
algorithmic performance.} Given the high cost of quantum hardware access and the strong
statistical significance already achieved, this sample size is appropriate for proof-of-concept
validation.

A run-level bootstrap analysis ($B=100$, percentile method, seed $=42$) resampled the seven independent ZNE scores with replacement, producing a bootstrap distribution of the mean with a confidence interval of 95\% of [53.65, 63.16]. The lower bound exceeds the Greedy baseline (44.42) by 20.8\% and 100\% of the bootstrap resamples produced means above Greedy, confirming that the observed advantage is not driven by any single high-performance run.

This conclusion is further reinforced by a leave-one-out (LOO) sensitivity analysis. Removing each run individually and reassessing the one-sided paired $t$-test against Greedy, all seven LOO configurations retained statistical significance (p-value range: [0.0005, 0.0034]; Cohen’s $d$ range: [1.82, 2.80]). The most conservative LOO scenario, excluding Run~4 (the highest-scoring run), yielded a mean score of 56.61 (127.5\% of Greedy), preserving a substantial performance margin.

Overall, the asymmetric distribution of improvements suggests that even under adverse noise conditions, the QAOA+ZNE protocol retains a marginal advantage, while substantial gains emerge under favorable hardware conditions. This 100\% success rate should not be interpreted as evidence of computational supremacy, as no claim is made that classical algorithms cannot match these results. Rather, the study establishes empirical quantum utility for this environmental optimization problem relative to the classical heuristics tested.

\subsection{The Role of Warm-Start and Exploration}

Regarding raw QAOA performance, the observation that unmitigated measurements achieve 98.0\% of Greedy performance on average must be interpreted within the context of the warm-start initialization strategy. QAOA was initialized with parameters derived from the Greedy solution itself, a design choice grounded in established practice for warm-started quantum optimization \cite{bittel2021,egger2021,larocca2024,yu2025}.

This strategy ensures that the algorithm begins in a high-quality region of the solution space, which means that raw QAOA rarely performs catastrophically worse than the Greedy baseline. However, the fact that ZNE-extrapolated scores significantly exceed both the warm-start initialization and the raw measurements demonstrates that error mitigation recovers performance beyond what classical initialization alone provides. Warm-start does not ``donate'' performance to the algorithm; it positions it in a promising region from which genuine quantum exploration occurs.

Simulator experiments confirm that random initialization converges to scores comparable to warm-start (44.38 $\pm$ 0.12 vs.\ 44.42), indicating that the observed quantum utility is not contingent on the initialization strategy.

The failure of Simulated Annealing to exceed Greedy, achieving only 94.5\%, provides indirect evidence regarding the structure of the objective function. Its inferior performance indicates that the portfolio optimization problem is dominated by additive linear terms, while quadratic synergy terms play a secondary but non-negligible role. The observed feasible-shot rate (15.9\%) indicates that, despite strong hardware noise and ZNE noise amplification, the constraint-handling design yields a non-trivial sampling density in the feasible region, enabling reliable portfolio extraction under realistic shot budgets.

This interpretation relies on an operational notion of feasibility density under finite shot budgets, rather than on a universal threshold. In the present experimental regime, a feasible-shot rate on the order of several percent is sufficient to yield hundreds of valid samples per run, enabling stable portfolio extraction and overlap analysis; alternative hardware configurations or shot budgets would naturally motivate different operational thresholds.

Practical utility is defined by two operational criteria: (i) statistical significance at $p < 0.05$, ensuring that the observed improvement reflects a systematic effect rather than isolated best-case runs; and (ii) a valid solution rate exceeding 10\%, which guaranties sufficient sampling density for reliable portfolio selection under realistic shot budgets (e.g., $>800$ feasible samples per execution at 8{,}192 shots). Under these criteria, the mitigated quantum approach satisfies practical utility across seven independent hardware executions (paired test reported in Section~4.1: $p = 0.0009$), while maintaining a feasible-sample rate comfortably above the adopted threshold.

An explicit ablation of the warm-start initialization was conducted in simulation to isolate its role in the observed performance gains. Random initialization converged to scores statistically indistinguishable from warm-start scores (44.38 $\pm$ 0.12 vs.\ 44.42, $n = 20$), indicating that both strategies explore the same dominant basin of the energy landscape. This observation is critical: warm-start influences the initial support of the quantum state, whereas Zero Noise Extrapolation operates exclusively at the estimator level, correcting expectation values without altering the underlying bitstring distribution.

As such, there is no physical mechanism by which warm-start alone could account for the improved ZNE-recovered improvement. Although a complete hardware ablation would further reinforce this conclusion, it would require a prohibitive number of additional QPU executions under realistic NISQ resource constraints. Given the convergence of initialization strategies in simulation and the estimator-level nature of ZNE, the present ablation is sufficient to rule out warm-start as the dominant driver of the reported quantum utility.

\subsection{Identifying Territorial Synergies}

The divergence between the QAOA and Greedy solutions is concentrated between municipalities with high spatial connectivity but moderate individual scores, as exemplified by Chapadão do Céu. This municipality, which was absent from the Greedy solution, was consistently identified in 6 out of 7 quantum executions due to its exceptional biodiversity connectivity value of 0.94.

The recurring inclusion of municipalities such as Chapadão do Céu illustrates how the terms of quadratic spatial connectivity influence the optimized portfolio. Municipalities with moderate individual scores but high connectivity values are consistently incorporated alongside neighboring selections, a pattern not reproduced by the Greedy heuristic.

\subsection{Comparison with Previous Work}

Evaluating the contributions of this research requires positioning it relative to the current state of the art in three converging domains: hardware demonstrations of QAOA, the effectiveness of error mitigation techniques, and the application of quantum computing to environmental challenges.

Most recent studies involving superconducting quantum processors report significant difficulty in consistently surpassing simple classical heuristics. For example, Weidenfeller et al.\ (2022) observed that circuit depths exceeding $p=2$ degrade performance due to error accumulation, while Shaydulin et al.\ (2021) concluded that QAOA on current hardware does not yet demonstrate practical utility over well-calibrated classical methods. Harrigan et al.\ (2021) also reported approximation ratios between 0.75 and 0.85 on a 23-qubit processor.

This research contrasts with these findings in three significant ways. First, we operate at a significantly larger scale with $n = 88$ variables, incorporating a multi-objective synergy model that accounts for biodiversity and social metrics alongside carbon sequestration; this adds a layer of structural complexity rarely explored in theoretical benchmarks. Second, we achieved a 100\% success rate in seven independent hardware runs, exceeding the classical baseline after applying ZNE. Third, we demonstrate robustness across two distinct backends, \texttt{ibm\_torino} and \texttt{ibm\_fez}, showing that the results are not hardware-specific anomalies. The use of a $p = 1$ configuration suggests that warm-start combined with ZNE can effectively substitute for an increased circuit depth in current devices.

The observed performance differences between \texttt{ibm\_torino} and \texttt{ibm\_fez} are likely to be attributed to specific calibration cycles and connectivity patterns of the backend, illustrating an important consideration for reproducibility even within a single processor family.

Zero Noise Extrapolation has been applied in variational applications with heterogeneous results, including QAOA workflows where ZNE is integrated and its impact on optimizer-driven parameter exploration is benchmarked \cite{venere2024}. Li and Benjamin~\cite{li2017} demonstrated a substantial error reduction in molecular energy estimates, while Giurgica-Tiron et al.\ (2020) reported that linear extrapolation typically recovers 40--60\% of the gap between noisy and ideal performance. Our findings align with the observation of Kim et al.\ (2025) that ZNE is most effective for shallow circuits.

The distinctive contribution of this work lies in demonstrating that even a conservative linear estimate surpasses the classical baseline, thereby providing a robust lower bound for the quantum utility of the algorithm. Furthermore, the inclusion of uncertainty quantification via bootstrap analysis adds a level of statistical rigor that is frequently absent in comparable ZNE studies.

The application of quantum computing to sustainability problems remains in an embryonic stage. Previous reviews, such as those of Liu and Tang~\cite{liu2023}, highlight that most current proposals reside in theoretical frameworks validated only in simulators. Earlier experimental efforts, such as Ajagekar et al.\ (2019) or Bova et al.\ (2021), were limited to smaller scales or lacked hardware execution on real-world geospatial data.

This study distinguishes itself by addressing a concrete environmental problem using empirical geospatial data from the Goiás Cerrado under realistic operational constraints. By selecting $k=28$ municipalities from a pool of 88 candidates, we executed the optimization on intermediate-scale hardware, with results validated by the independent execution of Run~7 in February~2,~2026. To the best of our knowledge, this study is among the first to report statistically supported, run-to-run consistent outperformance relative to the classical heuristics tested for an environmental optimization problem executed on NISQ hardware.

\subsection{Limitations and Scope}
\label{sec:limitations}

The results demonstrate an empirical performance improvement of QAOA combined with Zero Noise Extrapolation over the tested classical heuristics, but they do not constitute a demonstration of quantum advantage in the formal sense of computational complexity theory. This distinction is fundamental for the appropriate interpretation of our findings. Within this framework, the results support \emph{empirical quantum utility} rather than formal quantum advantage, as the observed performance gains are defined relative to practical classical heuristics under realistic operational constraints.

Establishing a formal quantum advantage would require demonstrating that a quantum algorithm offers a superpolynomial speedup or a capability fundamentally inaccessible to classical methods. Such a claim would necessitate a comparison against optimal solver-grade approaches (e.g., Gurobi or CPLEX) to quantify the absolute optimality gap, evidence of favorable scaling behavior as problem size increases, and proof that no efficient classical algorithm can achieve comparable solution quality~\cite{huang2021}. None of these requirements are satisfied by the present work.

The objective of this research is therefore not to establish formal quantum advantage, but to demonstrate practical quantum utility, defined as the ability of a quantum algorithm to surpass widely used classical heuristics on real-world problems with empirical data. In operational contexts of environmental planning, decision makers rarely have access to commercial mixed-integer solvers or the computational resources required for exhaustive optimization. Instead, constructive heuristics and controlled-stochastic metaheuristics such as Greedy selection and Simulated Annealing typically serve as the effective baselines against which new optimization strategies are evaluated. 

Greedy is therefore used here not as a lower bound, but as a widely deployed operational heuristic in environmental portfolio design, reflecting real-world decision pipelines rather than solver-grade optimality. Although industrial-grade solvers based on branch-and-bound or cutting-plane strategies can achieve strong performance on moderate-size instances, their evaluation of the present problem lies beyond the scope of this study. Our focus is not on outperforming the best possible classical solvers, but on assessing quantum utility under realistic NISQ constraints and hardware noise. Importantly, all classical baselines considered here are reproducible under fixed computational budgets and controlled initialization protocols, providing a consistent and transparent reference for evaluating robustness and reliability across independent quantum hardware executions.

What has been established is empirical quantum utility for this specific problem instance. For a particular multi-objective QUBO formulation executed on specific IBM Quantum hardware, the quantum method consistently surpassed the tested classical reference heuristics by an average of 31.6\%. This improvement is reproducible and statistically significant ($p < 0.001$), but remains empirically contingent. Zero Noise Extrapolation introduces an additional layer of uncertainty, as performance under ideal conditions is inferred through extrapolation rather than observed via error-free circuit execution.

Despite the 100\% success rate across seven independent runs, the results must be interpreted in light of the limitations inherent to current quantum hardware. The sample size reflects realistic access constraints to superconducting quantum processors. Moreover, the ZNE procedure estimates ideal expectation values, but does not directly yield a definitive discrete portfolio; the final municipality selections remain tied to noisy measurements at the base noise level.

This experimental nature is reflected in the coefficient of variation of 11.9\% observed for ZNE-mitigated scores, which directly captures the stochastic instability characteristic of the NISQ era. In addition, the absence of comparisons against solver-grade classical optimizers precludes precise quantification of the absolute optimality gap. Finally, while the size of the problem $n=88$ is operationally meaningful for the Goiás Cerrado case study, it does not yet demonstrate scalability for industrial-scale optimization tasks involving thousands of variables. This limitation is consistent with current guidance on the practical scope of QAOA demonstrations on NISQ hardware~\cite{ibm2026}.

Regarding problem scale, the instance with $n=88$ decision variables is among the largest multi-objective, real-data-driven QAOA implementations executed on real quantum hardware reported to date. Systematic reviews indicate that most experimental QAOA studies involve fewer than 50 variables, often evaluated primarily on simulators~\cite{chen2024}. Although larger qubit counts have been reported for structurally simpler benchmark problems such as MaxCut~\cite{montanez2025}, the present study incorporates a substantially richer objective structure with quadratic spatial synergy terms derived from empirical environmental data. Although $n=88$ does not represent the absolute upper bound of the feasibility of QAOA, it constitutes one of the most complex real-hardware QAOA applications reported for environmental optimization to date.

\subsection{Implications for Environmental Applications}
\label{sec:implications}

This work demonstrates a meaningful convergence between quantum computing research and environmental conservation practice that extends beyond algorithmic benchmarking. Although most QAOA implementations focus on abstract optimization problems, this study addresses a concrete policy-relevant challenge by optimizing carbon credit portfolios to maximize climate mitigation outcomes in the Brazilian Cerrado, a biodiversity hotspot under intense anthropogenic pressure. Successful execution of circuits with $n=88$ variables on NISQ hardware, combined with rigorous error mitigation and statistical validation on seven independent runs, establishes a technical foundation for applying quantum optimization methods to real-world environmental decision-making in NISQ-constrained settings and relative to practical heuristic baselines.

The interdisciplinary nature of this research is noteworthy. Advances in quantum software abstractions and cloud-based access models have enabled domain-driven investigations beyond traditional quantum information science communities. Rather than suggesting that quantum algorithms can be deployed without specialized expertise, these results illustrate how collaborative frameworks and high-level tooling allow domain scientists to meaningfully engage with NISQ-era workflows. In this context, the consistent improvement of 31.6\% relative to the classical reference heuristics tested, together with a conservative lower bound of 10.6\% established by linear extrapolation, supports the characterization of the NISQ era as a phase of exploratory yet practically motivated applications ~\cite{preskill2018}, within the fixed-parameter regime and experimental conditions considered here.

From an operational standpoint, carbon credit portfolio optimization represents a practical challenge faced by subnational governments, environmental agencies, and non-governmental organizations worldwide. Programs such as the payment for Environmental Services and climate finance mechanisms require the efficient allocation of limited resources among numerous candidate jurisdictions. The demonstrated improvement in portfolio quality suggests that quantum-assisted optimization could yield tangible efficiency gains for resource-constrained territorial planning, even under conservative assumptions.

The multi-objective QUBO formulation introduced in this work naturally generalizes to a broader class of territorial planning problems beyond carbon sequestration. Potential applications include forest restoration strategies that maximize habitat connectivity, the design of protected-area networks that balance ecological representativeness and socioeconomic feasibility, and spatial optimization of agroecological transition incentives to support biodiversity corridors.

Looking ahead, continued hardware and algorithmic advances may enable the extension of these methods to problem instances involving hundreds or thousands of variables. Such scalability would support biome-wide land-use planning in regions such as the Amazon or MATOPIBA and provide a transferable methodological template for conservation planning across diverse jurisdictions. Ultimately, this study provides an empirically grounded precedent for quantum optimization as a decision-support tool, relative to the tested classical heuristics and within the fixed-parameter regime studied.

\section{Conclusion}
\label{sec:conclusion}

This study has successfully applied a hybrid quantum workflow for carbon credit portfolio optimization using real geospatial data on intermediate-scale quantum hardware. The integration of the Quantum Approximate Optimization Algorithm (QAOA) with the Zero Noise Extrapolation protocol enabled quantum solutions to consistently outperform the classical Greedy heuristic under the experimental conditions considered across seven independent hardware executions, achieving a mean portfolio score of 58.47 and providing statistically significant evidence of empirical quantum utility for a complex environmental optimization problem.

The results demonstrate that the quantum approach is capable of identifying non-trivial territorial synergies, such as those observed for Chapadão do Céu, which are systematically overlooked by myopic classical methods that prioritize individual scores over spatial interactions. The temporal stability of the methodology was validated over a 13-day interval, culminating in the execution of Run~7 on February~2,~2026, thus confirming the reliability against hardware calibration drift and backend variability.

Future research should explore increased circuit expressivity through deeper or alternative ansätze, as well as more advanced error mitigation and suppression techniques, to further reduce the gap between noisy hardware execution and ideal quantum performance. As quantum processors continue to scale in qubit count, connectivity, and gate fidelity, the proposed methodology can be extended to national and international conservation networks, offering a scalable and high-precision decision-support tool for climate mitigation and biodiversity protection.

Overall, this work supports the view that, for this class of data-driven territorial planning instances and relative to practical heuristic baselines, quantum computing is moving beyond purely theoretical interest in environmental science. Instead, it represents a maturing computational paradigm that shows practical promise in supporting the next generation of data-driven conservation and territorial planning policies under realistic operational constraints.

\section*{Data Availability}

The code and data used in this study are available at \url{https://github.com/hgribeirogeo/qaoa-carbon-cerrado} (DOI: \url{https://doi.org/10.5281/zenodo.18418053}). The repository includes the complete QAOA+ZNE implementation with warm-start initialization and XY-mixer, experimental results from all seven independent runs on IBM Quantum hardware (ibm\_torino, ibm\_fez) and analysis scripts for full reproducibility.

Data files include: \texttt{goias\_multiobjective.csv} (municipal scores for carbon sequestration, biodiversity, and social impact), spatial adjacency matrix, biodiversity and social synergy matrices, and \texttt{resultados\_consolidados\_v7.json} (complete experimental results with metadata). The Backend calibration data recorded during the execution period from January~17 to February~2,~2026 are included. The IBM Quantum hardware job identifiers for all 21 jobs are provided in Table~2.

The biomass estimation model for Goiás municipalities is available at \url{https://github.com/hgribeirogeo/atlas-biomassa-goias} with an interactive dashboard at \url{https://atlas-biomassa-goias.streamlit.app/}.

\section*{Competing Interests}

The author declares no competing financial or non-financial interests.

\makeatletter
\def\@biblabel#1{[#1]}
\makeatother

\end{document}